# An enhanced statistical feature fusion approach using an improved distance evaluation algorithm and weighted K-nearest neighbor for bearing fault diagnosis

Amir Eshaghi Chaleshtori[a], Abdollah Aghaie[b,*]

[a]Ph.D. in Industrial Engineering, K.N.Toosi University of Technology, Tehran, Iran
[b]Professor of Industrial Engineering, K.N.Toosi University of Technology, Tehran, Iran

* Corresponding author: Abdollah Aghaie (aaghaie@kntu.ac.ir)

**Abstract**
Bearings are among the most failure-prone components in rotating machinery, and their condition directly impacts overall performance. Therefore, accurately diagnosing bearing faults is essential for ensuring system stability. However, detecting such malfunctions in noisy environments, where data is collected from multiple sensors, necessitates the extraction and selection of informative features. This paper proposes an improved distance evaluation algorithm combined with a weighted K-nearest neighbor (KNN) classifier for bearing fault diagnosis. The process begins with extracting and integrating statistical features of vibration across the time, frequency, and time-frequency domains. Next, the improved distance evaluation algorithm assigns weights to the extracted features, retaining only the most informative ones by eliminating insensitive features. Finally, the selected features are used to train the weighted KNN classifier. To validate the proposed method, we employ bearing data from the University of Ottawa. The results demonstrate the effectiveness of our approach in accurately identifying bearing faults.

*Keywords:* Bearing fault diagnosis; Feature fusion; Feature selection; Weighted KNN; Improved distance evaluation algorithm.

## 1. Introduction

Bearings, as critical components in rotating machinery, are highly susceptible to damage and failure [49, 60, 36, 17, 61]. Any bearing fault can degrade machine performance, cause unexpected breakdowns, and even lead to human fatalities [70, 6]. Predicting faults in advance helps prevent unforeseen damage to machine components, reducing maintenance costs and downtime [36, 38]. Therefore, accurately identifying bearing faults is essential [30, 47].

In bearing fault diagnosis, two primary approaches exist: physical models and data-driven models [28]. Physical models simulate system dynamics based on fundamental principles to analyze degradation. However, with increasing equipment complexity, understanding malfunctions and pinpointing their physical causes has become increasingly challenging. Consequently, data-driven models, which rely on historical data acquisition and processing, have proven to be more effective in fault prediction [3].

Despite their advantages, data-driven models face a key challenge: data collected in real-world working environments are often contaminated with environmental noise [19, 22, 22]. Thus, developing an effective feature engineering methodology that can identify the most valuable features while eliminating irrelevant ones is crucial for improving bearing fault diagnosis [60]. Various feature engineering methods exist for feature extraction and selection. Traditionally, filter-based [75], wrapper-based [29], and embedded methods [44] have been used to derive informative fault characteristics.

Several distance evaluation techniques are widely employed in feature engineering, including Euclidean distance (ED) [3], Mahalanobis distance (MD) [50, 51, 58], principal component analysis (PCA) [83, 64], independent component analysis (ICA) [34], canonical correlation analysis (CCA) [35], and linear discriminant analysis (LDA) [82]. However, conventional feature engineering methods primarily depend on historical data to assess the sensitivity of features. To address the limitations of these traditional techniques, deep learning has gained widespread application in bearing fault diagnosis, lifecycle assessment, and condition monitoring, owing to its powerful ability to extract nonlinear features and its high scalability [5, 78, 64, 12]. Several deep learning models are commonly used for bearing fault identification, including deep belief networks (DBN), autoencoders, artificial neural networks (ANN), and convolutional neural networks (CNN) [20, 71, 7, 33, 76].

This paper presents a novel feature fusion approach for identifying informative features in bearing fault diagnosis, particularly in scenarios where feature extraction is hindered by interference, such as noise. The proposed method first applies Fourier and discrete wavelet transforms to extract frequency and time-frequency content from bearing vibration data. Subsequently, statistical feature functions are employed across multiple domains including time, frequency, and time-frequency to generate high-dimensional feature representations. Next, a novel improved distance evaluation algorithm is applied to refine the feature set, ensuring the selection of the most informative features for fault detection. Finally, a weighted K-nearest neighbor (KNN) classifier is used for fault classification, and its accuracy is assessed.

To validate the proposed method, experiments are conducted using the University of Ottawa bearing dataset [21]. The primary contributions of this study are as follows:
- Development of a filter-based feature fusion algorithm to select informative features.
- Introduction of a novel weighting strategy that evaluates feature robustness and their discriminative ability for different bearing faults.
- Proposal of a new classifier for improved bearing fault classification.
- Demonstration of the proposed method's superiority over existing techniques.

The remainder of this paper is structured as follows: Section 2 provides a review of relevant research. Section 3 details the proposed fault diagnosis methodology, including data collection, feature extraction, and model development. Section 4 presents experimental validation using the Ottawa bearing dataset and evaluates the method's performance against existing approaches. Section 5 discusses the experimental results and comparative analysis. Finally, Section 6 concludes the study and outlines potential future research directions.

2. **Literature review**

Bearing failures can arise from various sources under operational conditions [41]. Therefore, diagnosing bearing faults effectively requires the collection and analysis of diverse data [10, 32]. Bearing fault diagnosis is commonly performed using vibration and acoustic data, with the primary distinction being whether the faulty bearing is directly in contact with the sensors [68]. However, the diagnostic process can be complicated by noise interference, which affects data accuracy and reliability [26, 4, 14, 15, 81]. Consequently, early and accurate detection of potential bearing failures is essential for ensuring system reliability.

Analyzing collected data and extracting relevant features can facilitate the identification of bearing faults [24, 2]. Time-domain analysis methods can provide diagnostic insights by computing statistical characteristics such as mean, skewness, kurtosis, coefficient of variation, root mean square (RMS), and variance [57, 74]. Additionally, several fault diagnosis techniques have been developed based on frequency-domain analysis, including Fourier spectroscopy, cepstrum analysis, and envelope spectrum analysis [72, 16, 39]. Furthermore, hybrid approaches such as wavelet analysis, short-time Fourier transform (STFT), and the Hilbert-Huang transform integrate both time- and frequency-domain information to enhance defect detection capabilities [73, 27, 66].

Despite these advancements, certain features may not be sensitive to early-stage faults [45], leading to a loss of critical bearing health information. Consequently, directly applying such features in fault diagnosis and prognosis may reduce model precision while increasing computational complexity. To address this issue, it is essential to extract informative data from high-dimensional feature sets by eliminating irrelevant or redundant information. Feature fusion techniques can improve diagnostic accuracy by integrating extracted features from multiple domains [77, 62, 69]. Feature fusion combines multiple original features to create a more informative, meaningful, and lower-dimensional feature representation [3, 79]. This approach offers significant advantages in detecting early-stage defects and classifying different fault types [8]. Several feature fusion methodologies have been explored in the literature. Traditionally, feature fusion is performed using filter-based, wrapper-based, or embedded methods [75, 44]. Principal component analysis (PCA) is one of the most widely used linear dimensionality reduction techniques, representing the maximum variance within the dataset [82, 64]. Independent component analysis (ICA) is another analytical technique employed for fault identification and noise filtering in data analysis [34]. Mahalanobis distance (MD) has been utilized for bearing health monitoring and detecting progressive damage during fault growth [58, 50, 51]. Additionally, canonical correlation analysis (CCA) has been applied in fault diagnosis; for example, Li et al. [35] used CCA to diagnose and identify faults in centrifugal compressors, while Zhou et al. [82] integrated CCA with a long short-term memory (LSTM) algorithm to assess system conditions following fault occurrence. While linear dimensionality reduction techniques can effectively analyze structured datasets, they may struggle to extract meaningful information from complex, nonlinear data. As a result, nonlinear dimensionality reduction methods have been introduced for real-world machine health monitoring applications [37]. Various nonlinear feature fusion techniques have been proposed for bearing fault diagnosis. Shao et al. [52] developed a local embedding method that applies linear transformations for dimensionality reduction. Wang et al. [67] introduced a technique based on Laplacian eigenmaps and spectral analysis to transform large-scale data into a more compact form. Sipola et al. [56] employed diffusion maps to maintain a high diffusion gap between data pairs, preserving the overall structure in a low-dimensional feature space. Su et al. [59] utilized the ISOMAP algorithm to extract significant feature spaces by considering local neighborhood information. Chen et al. [9] applied a local tangent space alignment method as a linear adaptation approach for dimensionality reduction. Zheng et al. [80] proposed a method based on conditional probability-based similarities to decrease data dimensionality.

A fundamental challenge in bearing fault diagnosis is selecting an appropriate feature fusion technique, as no unified framework currently exists for addressing this issue. This study proposes a novel feature fusion method for identifying features that improve bearing fault classification. Specifically, we introduce an improved distance evaluation algorithm as a new filter-based approach for selecting a subset of significant features. The selected feature subset is then input into a weighted K-nearest neighbor (KNN) classifier to detect and classify multiple fault types.

## 2. The suggested method for detecting bearing faults

Time-domain analysis is a fundamental technique for bearing fault diagnosis, as it provides information about a signal's amplitude over time. However, it does not capture frequency-related information. To address this limitation, frequency-domain analysis is employed to extract frequency-specific components, enabling the identification of characteristic defect frequencies. While frequency-domain analysis is effective in identifying different types of bearing defects, it lacks time-related information. Therefore, time-frequency techniques have been developed to simultaneously extract both time and frequency information, enhancing the effectiveness of bearing fault detection and prognosis.

In this study, vibration data is analyzed to monitor bearing conditions. First, statistical features are extracted from time-domain vibration data. Next, frequency-domain features are derived by applying the fast Fourier transform (FFT) to the vibration data. Additionally, discrete wavelet transform (DWT) is used to extract time-frequency domain features. As a result, a total of 112 features are identified across all three domains. The primary statistical characteristics are summarized in Table 1.

To perform signal processing, MATLAB® software is utilized. However, this high-dimensional feature set may contain redundant or overlapping characteristics, leading to overfitting, increased model complexity, and prolonged computation time. To mitigate these issues, a lower-dimensional subset of the most informative features must be selected. This study proposes an improved distance evaluation algorithm (PIDE) to achieve this objective. In PIDE, features are weighted based on their robustness against noise, their average intra-class distance, and their mean inter-class distance. A subset of the highest-ranked features is then selected.

This filter-based approach ensures that only the most significant features contribute to fault classification. To further enhance diagnostic accuracy, a novel weighted K-nearest neighbor (KNN) classifier is introduced. The weighted KNN model assigns feature weights derived from the improved distance evaluation algorithm, improving classification performance. The overall methodology is illustrated in Figure 1.

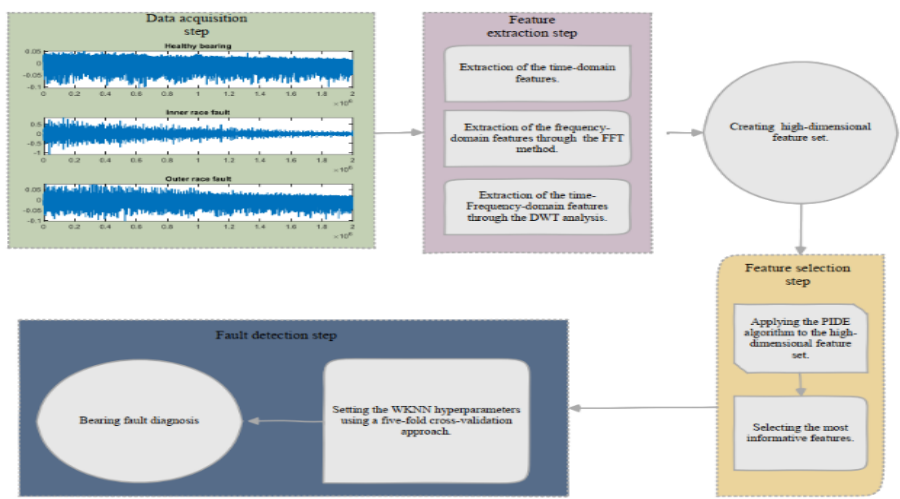

Figure 1. Flowchart of the proposed fault diagnosis approach

Table 1. Statistical features formulas.

| Feature | Description | Formula | Feature | Description | Formula |
|---|---|---|---|---|---|
| $F_1$ | Mean | $F_1 = \frac{\sum_{n=1}^{N} x(n)}{N}$ | $F_9$ | $3^{th}$ central moment | $F_9 = \frac{\sum_{n=1}^{N}(x(n)-F_1)^3}{(N-1)}$ |
| $F_2$ | Maximum | $F_2 = \max(x(n))$ | $F_{10}$ | $4^{th}$ central moment | $F_{10} = \frac{\sum_{n=1}^{N}(x(n)-F_1)^4}{(N-1)}$ |
| $F_3$ | Root mean square | $F_3 = \sqrt{\frac{1}{N}\sum_{n=1}^{N} x(n)^2}$ | $F_{11}$ | $5^{th}$ central moment | $F_{11} = \frac{\sum_{n=1}^{N}(x(n)-F_1)^5}{(N-1)}$ |
| $F_4$ | Standard deviation | $F_4 = \sqrt{\frac{1}{N-1}\sum_{n=1}^{N}(x(n)-F_1)^2}$ | $F_{12}$ | $6^{th}$ central moment | $F_{12} = \frac{\sum_{n=1}^{N}(x(n)-F_1)^6}{(N-1)}$ |
| $F_5$ | Impulse factor | $F_5 = \frac{\max(x(n))}{\frac{\sum_{n=1}^{N}|x(n)|}{N}}$ | $F_{13}$ | FM4 | $F_{13} = \frac{F_{10}}{F_4^2}$ |
| $F_6$ | Crest factor | $F_6 = \frac{\max(x(n))}{F_3}$ | $F_{14}$ | Variance | $F_{14} = F_4^2$ |
| $F_7$ | Skewness | $F_7 = \frac{\sum_{n=1}^{N}(x(n)-F_1)^3}{(N-1)F_4^3}$ | $F_{15}$ | Shape factor | $F_{15} = \frac{F_3}{\frac{\sum_{n=1}^{N}|x(n)|}{N}}$ |
| $F_8$ | Kurtosis | $F_8 = \frac{\sum_{n=1}^{N}(x(n)-F_1)^4}{(N-1)F_4^4}$ | $F_{16}$ | Entrpoy | $F_{16} = \sum_{n=1}^{N} x(n)\log\left(\frac{1}{x(n)}\right)$ |

3.1. *The process of acquiring data and extracting features*
The first step in diagnosing bearing faults using machine learning is feature extraction. This study utilizes statistical features to characterize vibration data across three domains: time, frequency, and time-frequency.
Time-domain analysis is a fundamental linear technique for examining data characteristics and structure. By analyzing signals collected from sensors, variations in time-domain features can be assessed. Since time-domain data preserves the most detailed information about bearing conditions, extracting features directly from raw data helps retain crucial diagnostic information, leading to an accurate representation of the bearing's fault state [41]. Table 1 summarizes the time-domain features extracted in this study. In the following formulas, N represents the number of data points, and $x(n)$ denotes the $n^{th}$ data point.
Beyond time-domain analysis, frequency-domain techniques are widely employed for bearing fault diagnosis. These methods transform time-domain signals into their frequency components, facilitating the identification of characteristic defect frequencies. Fast Fourier Transform (FFT) is a commonly used approach for this purpose [25]. Compared to time-domain analysis, frequency-domain analysis provides additional diagnostic insights [63]. Table 1 presents the frequency-domain features extracted in this study using the FFT method.
However, frequency-domain methods have certain limitations. Due to harmonic effects and fault frequencies, they often struggle to accurately identify fault peaks [43]. Moreover, FFT assumes signal periodicity, making it less suitable for analyzing non-stationary data. Given that operating conditions frequently fluctuate, bearing vibration signals are typically non-stationary [46].
To overcome the challenges associated with nonlinear and non-stationary data, time-frequency analysis offers a more comprehensive representation by examining the relationship between time and frequency [13]. This study employs discrete wavelet transform (DWT), a robust signal processing tool with excellent time-frequency localization properties [53].
DWT decomposes signals using high-pass and low-pass filters. High-pass filters preserve high-frequency components (i.e., details), while low-pass filters retain low-frequency components (i.e., approximations). The number of filter applications depends on the decomposition level. For a given mother wavelet ψ(t), DWT is defined as:

$$\psi_{j,k}(t) = \frac{1}{\sqrt{2^j}}\psi\left(\frac{t - 2^j k}{2^j}\right) \tag{1}$$

where *j* and *k* denote the scale and translation indices, respectively [54]. In this study, Daubechies 10 (db10) wavelet with four decomposition levels is used as the mother wavelet to decompose vibration data [55]. The extracted time-frequency features under various bearing conditions are summarized in Table 1.

3.2. *The improved distance evaluation algorithm*
   Feature fusion extends the feature set, providing a more comprehensive signal representation. However, increasing dataset size introduces several challenges:1) Larger datasets increase model complexity, making training more time-intensive. 2) If the selected features lack significant variation across classes, classification accuracy may suffer. These uninformative features are insensitive to classification and must be eliminated to prevent negative impacts.
The conventional distance evaluation algorithm [31] measures intra-class and inter-class distances by treating each data group as a whole. It selects informative features by identifying those with small intra-class distances and large inter-class distances. However, a major drawback of this approach is its sensitivity to noise, which can distort the distance metrics. To overcome this limitation, this study introduces an improved distance evaluation algorithm that enhances robustness against noise and outliers. This novel approach evaluates feature cohesion through robustness analysis—a measure of how tightly data points cluster around a central value. Compared to conventional methods, this technique provides more information on data dispersion and latent differences across the dataset. The robustness metric helps assess a feature's resistance to random deviations caused by noise interference, stochastic bearing degradation, or operational fluctuations.

Given a $J$-dimensional feature set with $S$ classes, where $N_s$ represents the number of samples per class ($J$, $S$, and $N_s$ are positive integers), the proposed improved distance evaluation algorithm is formulated as follows:

$$\{q_{n,s,j};\ s = 1,2,\cdots,S;\ n = 1,2,\cdots,N_s;\ j = 1,2,\cdots,J\} \quad (2)$$

where $q_{n,s,j}$ denotes the $j^{th}$ feature of the $n^{th}$ sample in the $s^{th}$ class.

In Algorithm 1, the first two steps represent the intra-class distance correlation, in which smaller average intra-class distances are more effective for classifying. Steps 3 ~ 4 illustrate the relationship between inter-class distances and a greater average distance between categories, resulting in better classification. The sensitivity is measured in steps 5 ~ 6 using the robustness metric, in which $N_s$ is the total number of feature values, and $x_j^n$ th is the average trend of the $j^{th}$ feature at the $n^{th}$ sample, typically obtained through smoothing. Steps 7 ~ 8 assign a weight to each feature for selecting informative ones. Then, the features with weights higher than the specified threshold are selected.

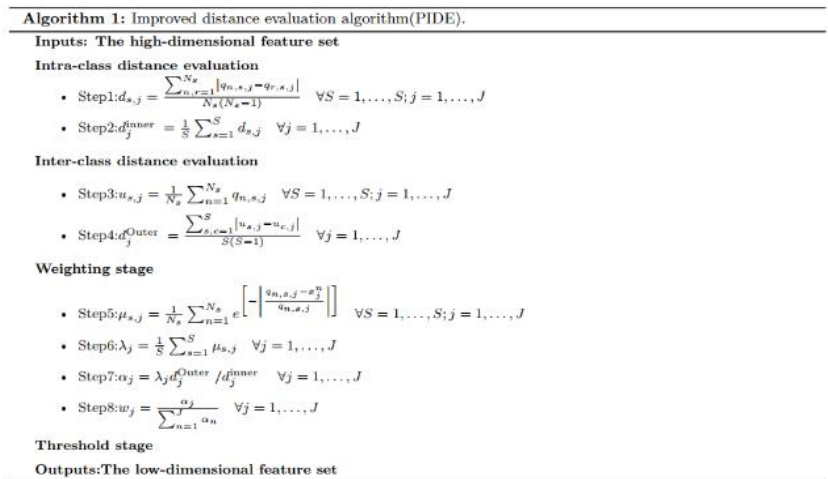

Figure 2. The proposed PIDE algorithm

3.3. *Weighted KNN classifier*

The weighted KNN method is an enhanced version of the KNN model. As a result, it not only offers the advantages of high computation efficiency when compared with KNN but also avoids the limitation that the classification accuracy of KNN is susceptible to the neighborhood size (k) [15]. This paper employs a weighted Euclidean distance based on the weights derived from the PIDE algorithm. Hence, the following Equation that utilizes the weights calculated by Algorithm 1 is employed in the weighted KNN method:

$$h_i = \left| \sum w_j(x_{ii} - y_{ii})^2 \right| \quad \forall i = 1,2,\ldots,N_s \quad (3)$$

In Equation (3), $w_j$ indicates the weight of $j^{th}$ feature obtained by Algorithm 1, and $N_s$ illustrates the sample size. Moreover, $h_i$ indicates the weighted distance between the unknown sample ($x_{ij}$) and the training sample ($y_{ij}$). As a result, the detection ability is enhanced, and the susceptibility to the neighborhood size (k) is considerably reduced, ultimately improving classification performance and robustness.

## 4. Experiments

### 4.1. Dataset for experiments

The experiments are conducted using bearing vibration data collected at the University of Ottawa [21]. A SpectraQuest Machinery Fault Simulator (MFS-PK5M) is used for testing. Figure 3 illustrates the experimental setup. As shown, a motor drives the shaft, with an AC drive regulating its rotational speed. Two ER16K ball bearings support a healthy shaft, while experimental bearings in various health states are tested by replacing them accordingly.

The experimental bearing is equipped with an ICP accelerometer (model 623C01) to collect vibration data. Additionally, an incremental encoder (EPC model 775) measures the shaft's rotational speed. To ensure data reliability, three trials are conducted for each experimental condition.

Table 2 presents the fault states analyzed in this study along with their corresponding labels. Vibration data is sampled at a rate of 200 kHz, with each state labeled in real-time as it occurs. The dataset also includes additional research data not described here, as it is not used in this study.

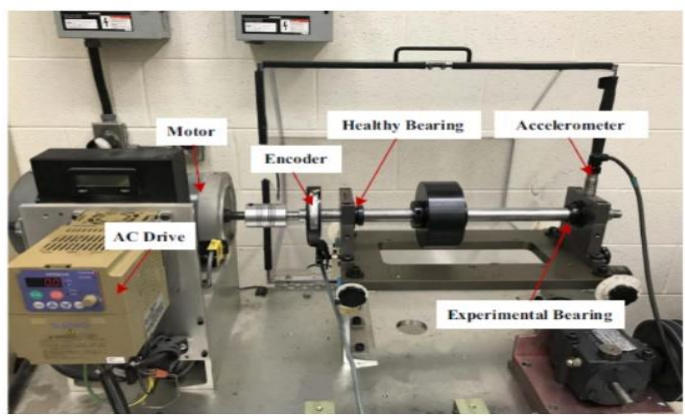

Figure 3 University of Ottawa time-varying bearing experiment.

*Table 2. Dataset overview showing each bearing state and its corresponding label.*

| Status | Speed varying conditions | label |
|---|---|---|
| Healthy bearing | | 1 |
| Inner race fault | Decreasing speed | 2 |
| Outer race fault | | 3 |

### 4.2. Overview of evaluation criteria

This study evaluates the effectiveness of the proposed method using standard classification performance metrics, presenting the results through a confusion matrix and a Receiver Operating Characteristic (ROC) curve.

In a typical classification problem, the class of interest is considered positive, while all other classes are treated as negative. After analyzing the dataset, the classifier produces four possible prediction outcomes:

- True Positive (TP): Positive classes correctly predicted as positive.
- False Positive (FP): Negative classes incorrectly predicted as positive.
- True Negative (TN): Negative classes correctly predicted as negative.
- False Negative (FN): Positive classes incorrectly predicted as negative.

The classification accuracy is generally defined as follows:

$$Accuracy = \frac{TP + TN}{TP + TN + FN + FP} \quad (4)$$

We also evaluate our method by computing the area under the curve (AUC) using a ROC curve. Generally, AUC ranges from 0.5 to 1 (i.e., random prediction to perfect prediction). Therefore, the area under the ROC curve indicates the overall accuracy of the method as it relates to how precisely faults and faultless groups are separated.

4.3. *Procedure for experiments*

A comparative analysis is conducted between the proposed method (PIDE-WKNN) and several general approaches, including principal component analysis KNN (PCA-KNN) proposed by Ahmad et al. [1], principal component analysis SVM (PCA-SVM) by Han et al. [18], linear discriminant analysis KNN (LDA-KNN) by Mehta et al. [42], linear discriminant analysis SVM (LDA-SVM) by Saha et al. [48], conventional improved distance evaluation SVM (CIDE-SVM), and conventional improved distance evaluation KNN (CIDE-KNN) proposed by Lei et al. [31].

Hyperparameter tuning is essential for both SVM and KNN classifiers. To optimize their performance and mitigate the effects of dataset variability, fivefold cross-validation is employed. The model integrates multi-domain features, with 80% of the dataset used for training and the remaining 20% for testing.

The fivefold cross-validation process is structured as follows: the dataset is first divided into training and testing sets. Four folds are used for training, while the remaining fold is reserved for testing. This process is repeated five times, ensuring each fold serves as a test set once. The parameters that yield the best results across all training iterations are selected for final classification. Notably, the number of selected features remains constant throughout the entire process. Table 3 presents the hyperparameters used in the experiment.

Table 3. Accuracy of comparison methods

| Methods | SVM hyperparameters | | | KNN hyperparameters | | | # Selected Features | Accuracy(%) |
|---|---|---|---|---|---|---|---|---|
| | Coding | Feasible point box-constraint | Kernel-scale | Coding | # neighbors | Distance | | |
| LDA − SVM | one-vs-one | 998.96 | 0.002188 | - | - | - | 112 | 96.67 |
| PCA − SVM | one-vs-one | 749.2 | 408.08 | - | - | - | 112 | 30 |
| CIDE − SVM | one-vs-all | 586.26 | 422.57 | - | - | - | 109 | 50 |
| PIDE − SVM | one-vs-one | 0.0048274 | 0.093014 | - | - | - | 21 | 90 |
| LDA − KNN | - | - | - | one-vs-all | 1 | seuclidean | 112 | 100 |
| PCA − KNN | - | - | - | one-vs-all | 1 | spearman | 112 | 96.67 |
| CIDE − KNN | - | - | - | one-vs-all | 1 | seuclidean | 109 | 100 |
| **PIDE − WKNN** | - | - | - | **one-vs-all** | **2** | **Weighted Euclidean** | **21** | **100** |

5. **Analysis and discussion of results**

5.1. *Results*

The proposed method is rigorously evaluated against other approaches in terms of classification accuracy using fivefold cross-validation. Table 3 summarizes the results, highlighting the best-performing methods. As shown in Table 3, the proposed method outperforms others, achieving classification accuracies of 96.67% (LDA-SVM), 30% (PCA-SVM), 50% (CIDE-SVM), 90% (PIDE-SVM), 100% (LDA-KNN), 96.67% (PCA-KNN), 100% (CIDE-KNN), and 100% (PIDE-WKNN).

A comparative analysis confirms the proposed method's effectiveness in fault identification, demonstrating the highest accuracy among all evaluated approaches. Furthermore, the results indicate that the proposed method achieves an average AUC of 1 across the three fault types, highlighting its optimal performance and high-accuracy fault diagnosis.

5.2. *Discussions*

As shown in Figure 4, time-frequency domain features exhibit a strong internal correlation, whereas frequency domain features demonstrate a lower internal correlation. Additionally, fused features help reduce feature correlation and eliminate redundant components, making them more effective for detecting informative features. This, in turn, positively impacts bearing fault prediction.

To evaluate the performance of different feature sets in bearing fault classification, we analyze the results of KNN classifiers trained using features from the time domain, frequency domain, time-frequency domain, and fused domain. For individual feature sets, the results indicate that time-frequency and time-domain features outperform frequency-domain features. However, fused features extracted using the proposed method achieve the highest classification accuracy.

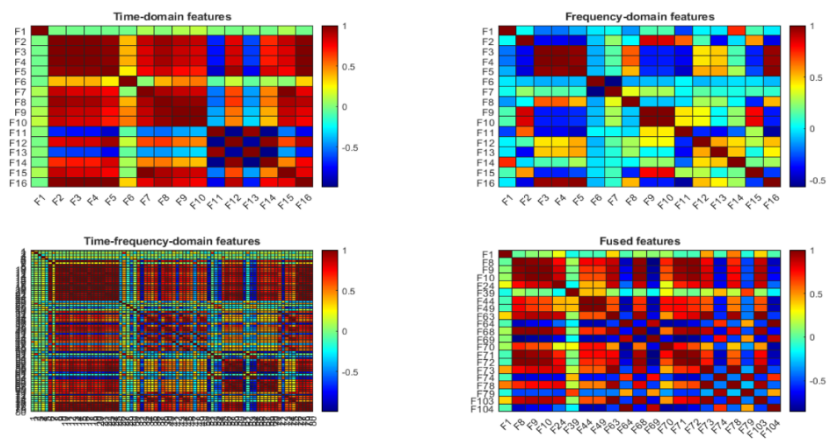

Figure 4. Correlation analysis of feature sets derived from various domains.

## 5.3. Analyzing the parameter of proposed fault diagnosis method

Although the proposed method demonstrates excellent diagnostic accuracy, its performance depends on the threshold level set in the PIDE algorithm for selecting informative features. To assess this impact, we generate different threshold scenarios and evaluate their effect on PIDE-WKNN performance. The relationship between threshold level and classification accuracy is illustrated in Figure 5.

As shown in Figure 5, the number of selected features decreases as the threshold level increases. Notably, raising the threshold level up to 0.92 does not reduce classification accuracy, indicating that a simpler yet highly accurate fault diagnosis classifier can be trained. However, when the threshold exceeds 0.92, both the number of selected features and the model's accuracy decline. Therefore, the results suggest that a threshold level of 0.92 yields the highest overall performance for the proposed method.

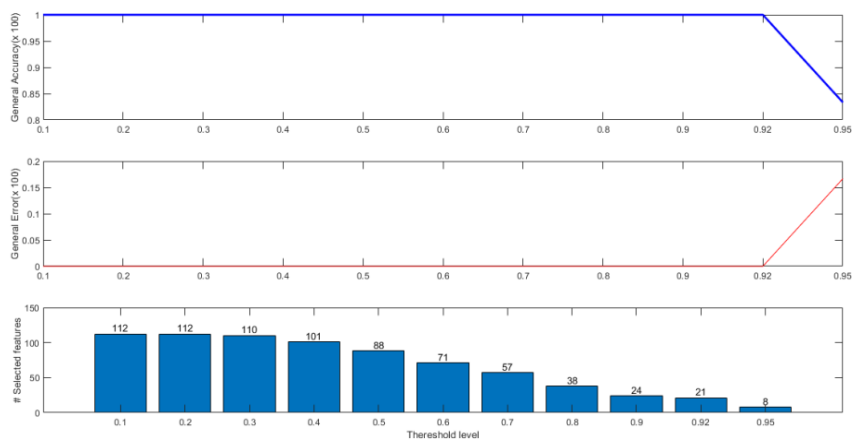

Figure 5. The accuracy results of PIDE-KNN method and the number of selected features under a variety of threshold scenarios.

## 5.4. Evaluation of the robustness of the suggested method

Since noisy data cannot be identified in practice, artificial noise is introduced in this experimental analysis. This study employs Algorithm 2, proposed by Ref. [40], to generate noisy data. To evaluate the robustness of the proposed method (PIDE), we compare it with other techniques, including principal component analysis (PCA), linear discriminant analysis (LDA), and the conventional distance evaluation method (CIDE) introduced by Lei et al. [31].

Robustness is assessed based on a metric that quantifies a feature's ability to withstand random fluctuations in operating conditions, such as sensor noise, bearing degradation, or temperature variations [11]. The robustness measure is defined by the following equation:

$$Rob(x) = \frac{1}{k} \sum \exp\left(-\left|\frac{x_k - x_k^T}{\cdot}\right|\right) \quad (5)$$

In Equation (5), K represents the total number of feature values, while $x_k$ denotes the feature value at time $t_k$. The term $x_k^T$ refers to the mean trend of the feature at time $t_k$, typically determined using a smoothing method.

Figure 7 illustrates the average and standard deviation of various bearing states under different noise ratios, ranging from 0.05 to 0.5. As shown in Figure 7(a), the proposed algorithm and LDA exhibit greater robustness compared to other methods as the noise ratio increases. Figure 7(b) further demonstrates that the proposed method produces more stable results than the alternatives.

To comprehensively validate the performance metrics, we conduct Kruskal-Wallis and Wilcoxon signed-rank tests to determine which algorithm outperforms the others [23]. The results, summarized in Tables 4 and 5, indicate that the comparison methods vary in terms of the mean and standard deviation of the robustness metric. These findings confirm the superiority of the proposed method over other approaches. However, the red values in Table 5 indicate cases where the specified metric does not show a statistically significant difference between two algorithms. Despite the inherent uncertainties of traditional methods, the proposed approach consistently outperforms competing algorithms.

**Algorithm 2: Generate Noisy data.**
**Inputs:** Features vector, InstanceRatio, FeatureRatio, sample size($N_s$), and the number of features($J$).
1. Randomly select $[InstanceRatio * N_s]$ instances.
2. Randomly select $[FeatureRatio * J]$ features.
   For each selected feature with the mean=$\mu$ and the standard deviation=$\sigma$)
3. Each entry of these selected instances is assigned with a random value within the range in ($\mu \pm 2 * \sigma$).

**Outputs:** A noisy feature subset

Figure 6. Generating noisy data algorithm

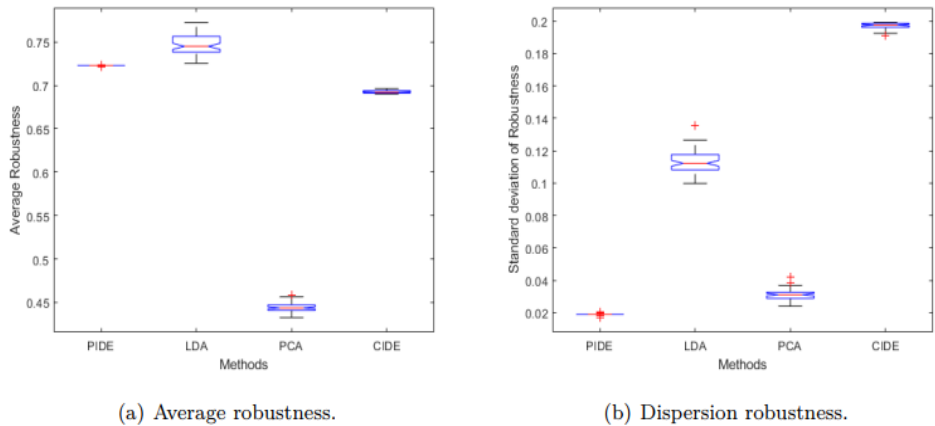

(a) Average robustness.  (b) Dispersion robustness.

Figure 7. The average and dispersion of the robustness values for comparison algorithms.

Table 4. The results of the non-parametric Kruskal-Wallis test.

| Performance metric | Test statistics | | |
|---|---|---|---|
| | $K - W\, statistics$ | df | Sig. |
| Mean | 240.66 | 3 | 6.8486e-52 |
| Standard deviation | 240.72 | 3 | 6.4656e-52 |

Table 5. The results of Wilcoxon signed-rank test.

| Wilcoxon signed-rank Statistics | Performance metric | Test Pairs | | |
|---|---|---|---|---|
| | | PIDE vs. CIDE. | PIDE vs. PCA. | PIDE vs. LDA. |
| Z | Mean | 10.0260 | 10.0260 | -10.0260 |
| Sig. | | 1.1723e-23 | 1.1723e-23 | 1.1723e-23 |
| Z | Standard deviation | -10.0260 | -10.0260 | -10.0260 |
| Sig. | | 1.1723e-23 | 1.1723e-23 | 1.1723e-23 |

## 6. Conclusions and prospects

This study proposes a novel bearing fault diagnosis method that integrates a weighted KNN classifier with an improved evaluation distance algorithm (PIDE). By extracting features from the time, frequency, and time-frequency domains, the method establishes a high-dimensional feature set. To address the challenges posed by high dimensionality, the PIDE algorithm is employed for feature selection. Subsequently, the weighted KNN classifier is used to determine the operational states of the bearing.

The proposed diagnostic model is tested and validated using vibration data from a real-world case study. Experimental results demonstrate that the method reliably detects bearing malfunctions with high accuracy. Furthermore, findings indicate that the model exhibits strong generalizability, effectively identifying bearing faults under diverse test conditions.

As a result, this model can be applied across various operational scenarios, including different working conditions for the same device, varying environments for different machines, and challenging industrial applications. Its practical value extends to predictive maintenance, enabling timely coordination of maintenance services.

Future research will focus on analyzing various fault types with different severity levels and investigating fault diagnosis across different life cycle stages. Additionally, further studies will explore ways to enhance the proposed method's robustness and generalization capabilities under continuously changing operating conditions.

**References**
[1] Ahmad, Z., Nguyen, T.K., Ahmad, S., Nguyen, C.D., Kim, J.M., 2022. Multistage centrifugal pump fault diagnosis using informative ratio principal component analysis. Sensors 22, 179.
[2] Behera, S., Misra, R., 2023. A multi-model data-fusion based deep transfer learning for improved remaining useful life estimation for iiot based systems. Engineering Applications of Artificial Intelligence 119, 105712.
[3] Buchaiah, S., Shakya, P., 2022. Bearing fault diagnosis and prognosis using data fusion-based feature extraction and feature selection. Measurement 188, 110506.
[4] Caesarendra, W., Pratama, M., Kosasih, B., Tjahjowidodo, T., Glowacz, A., 2018. Parsimonious network based on a fuzzy inference system (panfis) for time series feature prediction of low-speed slew bearing prognosis. Applied Sciences 8, 2656.
[5] Cai, B., Shao, X., Liu, Y., Kong, X., Wang, H., Xu, H., Ge, W., 2019. Remaining useful life estimation of structure systems under the influence of multiple causes: Subsea pipelines as a case study. IEEE Transactions on Industrial Electronics 67, 5737–5747.
[6] Chaleshtori, A.E., et al., 2022. Data fusion techniques for fault diagnosis of industrial machines: a survey. arXiv preprint arXiv:2211.09551.
[7] Che, C., Wang, H., Ni, X., Fu, Q., 2020. Domain adaptive deep belief network for rolling bearing fault diagnosis. Computers & Industrial Engineering 143, 106427.
[8] Che, C., Wang, H., Ni, X., Lin, R., 2021. Hybrid multimodal fusion with deep learning for rolling bearing fault diagnosis. Measurement 173, 108655.
[9] Chen, F., Tang, B., Song, T., Li, L., 2014. Multi-fault diagnosis study on roller bearing based on multi-kernel support vector machine with chaotic particle swarm optimization. Measurement 47, 576–590.
[10] Chen, X., Wang, S., Qiao, B., Chen, Q., 2018. Basic research on machinery fault diagnostics: Past, present, and future trends. Frontiers of Mechanical Engineering 13, 264–291.
[11] Duong, B.P., Khan, S.A., Shon, D., Im, K., Park, J., Lim, D.S., Jang, B., Kim, J.M., 2018. A reliable health indicator for fault prognosis of bearings. Sensors 18, 3740.
[12] Gai, J., Shen, J., Hu, Y., Wang, H., 2020. An integrated method based on hybrid grey wolf optimizer improved variational mode decomposition and deep neural network for fault diagnosis of rolling bearing. Measurement 162, 107901.
[13] Gao, Y., Yu, D., Wang, H., 2020. Fault diagnosis of rolling bearings using weighted horizontal visibility graph and graph fourier transform. Measurement 149, 107036.
[14] Glowacz, A., 2018. Recognition of acoustic signals of commutator motors. Applied Sciences 8, 2630.
[15] Glowacz, A., 2019. Fault detection of electric impact drills and coffee grinders using acoustic signals. Sensors 19, 269.
[16] Gu, X., Yang, S., Liu, Y., Hao, R., 2016. Rolling element bearing faults diagnosis based on kurtogram and frequency domain correlated kurtosis. Measurement Science and Technology 27, 125019.

[17] Haidong, S., Hongkai, J., Xingqiu, L., Shuaipeng, W., 2018. Intelligent fault diagnosis of rolling bearing using deep wavelet auto-encoder with extreme learning machine. Knowledge-Based Systems 140, 1–14.
[18] Han, T., Zhang, L., Yin, Z., Tan, A.C., 2021. Rolling bearing fault diagnosis with combined convolutional neural networks and support vector machine. Measurement 177, 109022.
[19] He, Y., Wang, X., Chen, Y., Du, Z., Huang, W., Chai, X., 2013. A simulation cloud monitoring framework and its evaluation model. Simulation Modelling Practice and Theory 38, 20–37.
[20] Hoang, D.T., Kang, H.J., 2019. A survey on deep learning-based bearing fault diagnosis. Neurocomputing 335, 327–335.
[21] Huang, H., Baddour, N., 2018. Bearing vibration data collected under time-varying rotational speed conditions. Data in brief 21, 1745–1749.
[22] Huang, M., Liu, Z., Tao, Y., 2020. Mechanical fault diagnosis and prediction in iot based on multi-source sensing data fusion. Simulation Modelling Practice and Theory 102, 101981.
[23] Jahani, H., Chaleshtori, A.E., Khaksar, S.M.S., Aghaie, A., Sheu, J.B., 2022. Covid-19 vaccine distribution planning using a congested queuing system—a real case from australia. Transportation Research Part E: Logistics and Transportation Review 163, 102749.
[24] Jardine, A.K., Lin, D., Banjevic, D., 2006. A review on machinery diagnostics and prognostics implementing condition-based maintenance. Mechanical systems and signal processing 20, 1483–1510.
[25] Javed, K., Gouriveau, R., Zerhouni, N., Nectoux, P., 2014. Enabling health monitoring approach based on vibration data for accurate prognostics. IEEE Transactions on industrial electronics 62, 647–656.
[26] Jiang, Y., Li, Z., Zhang, C., Hu, C., Peng, Z., 2016a. On the bi-dimensional variational decomposition applied to nonstationary vibration signals for rolling bearing crack detection in coal cutters. Measurement Science and Technology 27, 065103.
[27] Jiang, Y., Zhu, H., Li, Z., 2016b. A new compound faults detection method for rolling bearings based on empirical wavelet transform and chaotic oscillator. Chaos, Solitons & Fractals 89, 8–19.
[28] Jung, D., Ng, K.Y., Frisk, E., Krysander, M., 2018. Combining model-based diagnosis and data-driven anomaly classifiers for fault isolation. Control Engineering Practice 80, 146–156.
[29] Kohavi, R., John, G.H., 1997. Wrappers for feature subset selection. Artificial intelligence 97, 273–324.
[30] Lei, Y., He, Z., Zi, Y., 2008a. Application of a novel hybrid intelligent method to compound fault diagnosis of locomotive roller bearings. Journal of vibration and acoustics 130.
[31] Lei, Y., He, Z., Zi, Y., 2008b. Fault diagnosis based on novel hybrid intelligent model. Chinese journal of mechanical engineering 44, 112–117.
[32] Lei, Y., Yang, B., Jiang, X., Jia, F., Li, N., Nandi, A.K., 2020. Applications of machine learning to machine fault diagnosis: A review and roadmap. Mechanical Systems and Signal Processing 138, 106587.
[33] Li, C., Bao, Z., Li, L., Zhao, Z., 2020. Exploring temporal representations by leveraging attention-based bidirectional lstm-rnns for multi-modal emotion recognition. Information Processing & Management 57, 102185.
[34] Li, J., Meng, Z., Yin, N., Pan, Z., Cao, L., Fan, F., 2021. Multi-source feature extraction of rolling bearing compression measurement signal based on independent component analysis. Measurement 172, 108908.
[35] Li, X., Duan, F., Loukopoulos, P., Bennett, I., Mba, D., 2018. Canonical variable analysis and long short-term memory for fault diagnosis and performance estimation of a centrifugal compressor. Control Engineering Practice 72, 177–191.
[36] Liang, M., Faghidi, H., 2014. Intelligent bearing fault detection by enhanced energy operator. Expert systems with applications 41, 7223–7234.
[37] Liang, P., Wang, B., Jiang, G., Li, N., Zhang, L., 2023. Unsupervised fault diagnosis of wind turbine bearing via a deep residual deformable convolution network based on subdomain adaptation under time-varying speeds. Engineering Applications of Artificial Intelligence 118, 105656.
[38] Liang, P., Wang, W., Yuan, X., Liu, S., Zhang, L., Cheng, Y., 2022. Intelligent fault diagnosis of rolling bearing based on wavelet transform and improved resnet under noisy labels and environment. Engineering Applications of Artificial Intelligence 115, 105269.
[39] Liao, Y., Sun, P., Wang, B., Qu, L., 2018. Extraction of repetitive transients with frequency domain multipoint kurtosis for bearing fault diagnosis. Measurement Science and Technology 29, 055012.
[40] Lin, C.C., Kang, J.R., Liang, Y.L., Kuo, C.C., 2021. Simultaneous feature and instance selection in big noisy data using memetic variable neighborhood search. Applied Soft Computing 112, 107855.
[41] Mao, X., Zhang, F., Wang, G., Chu, Y., Yuan, K., 2021. Semi-random subspace with bi-gru: Fusing statistical and deep representation features for bearing fault diagnosis. Measurement 173, 108603.
[42] Mehta, A., Goyal, D., Choudhary, A., Pabla, B., Belghith, S., 2021. Machine learning-based fault diagnosis of self-aligning bearings for rotating machinery using infrared thermography. Mathematical Problems in Engineering 2021.
[43] Ocak, H., Loparo, K.A., Discenzo, F.M., 2007. Online tracking of bearing wear using wavelet packet decomposition and probabilistic modeling: A method for bearing prognostics. Journal of sound and vibration 302, 951–961.
[44] Osman, H., Ghafari, M., Nierstrasz, O., 2017. Automatic feature selection by regularization to improve bug prediction accuracy, in: 2017 IEEE Workshop on Machine Learning Techniques for Software Quality Evaluation (MaLTeSQuE), IEEE. pp. 27–32.
[45] Pan, L., Zhao, L., Song, A., She, S., Wang, S., 2021. Research on gear fault diagnosis based on feature fusion optimization and improved two hidden layer extreme learning machine. Measurement 177, 109317.
[46] Peng, Z.K., Chu, F., 2004. Application of the wavelet transform in machine condition monitoring and fault diagnostics: a review with bibliography. Mechanical systems and signal processing 18, 199–221.
[47] Qian, W., Li, S., Lu, J., 2022. Adaptive nearest neighbor reconstruction with deep contractive sparse filtering for fault diagnosis of roller bearings. Engineering Applications of Artificial Intelligence 111, 104749.
[48] Saha, D.K., Hoque, M.E., Badihi, H., 2022. Development of intelligent fault diagnosis technique of rotary machine element bearing: A machine learning approach. Sensors 22, 1073.
[49] Saimurugan, M., Ramachandran, K., Sugumaran, V., Sakthivel, N., 2011. Multi component fault diagnosis of rotational mechanical system based on decision tree and support vector machine. Expert Systems with Applications 38, 3819–3826.
[50] Shakya, P., Darpe, A., Kulkarni, M., Sas, P., Moens, D., Jonckheere, S., 2012. Use of mahalanobis taguchi system as data fusion approach for monitoring health of rolling element bearing, in: ISMA International Conference on Noise and Vibration Engineering, pp. 3389–3402.

[51] Shakya, P., Kulkarni, M.S., Darpe, A.K., 2014. A novel methodology for online detection of bearing health status for naturally progressing defect. Journal of Sound and Vibration 333, 5614–5629.
[52] Shao, H., Jiang, H., Li, X., Liang, T., 2018. Rolling bearing fault detection using continuous deep belief network with locally linear embedding. Computers in Industry 96, 27–39.
[53] Shibata, K., Takahashi, A., Shirai, T., 2000. Fault diagnosis of rotating machinery through visualisation of sound signals. Mechanical Systems and Signal Processing 14, 229–241.
[54] Shukla, K.K., Tiwari, A.K., 2013. Efficient algorithms for discrete wavelet transform: with applications to denoising and fuzzy inference systems. Springer Science & Business Media.
[55] Shukla, R., Kankar, P., Pachori, R., 2021. Automated bearing fault classification based on discrete wavelet transform method. Life Cycle Reliability and Safety Engineering 10, 99–111.
[56] Sipola, T., Ristaniemi, T., Averbuch, A., 2015. Gear classification and fault detection using a diffusion map framework. Pattern Recognition Letters 53, 53–61.
[57] Song, J., Zhao, J., Zhang, X., Dong, F., Zhao, J., Xu, L., Yao, Z., 2019. Accurate demagnetization faults detection of dualsided permanent magnet linear motor using enveloping and time-domain energy analysis. IEEE Transactions on Industrial Informatics 16, 6334–6346.
[58] Soylemezoglu, A., Jagannathan, S., Saygin, C., 2010. Mahalanobis taguchi system (mts) as a prognostics tool for rolling element bearing failures. Journal of Manufacturing Science and Engineering 132.
[59] Su, H., Xiang, L., Hu, A., Gao, B., Yang, X., 2021. A novel hybrid method based on kelm with sapso for fault diagnosis of rolling bearing under variable operating conditions. Measurement 177, 109276.
[60] Sugumaran, V., Ramachandran, K., 2011. Effect of number of features on classification of roller bearing faults using svm and psvm. Expert Systems with Applications 38, 4088–4096.
[61] Sun, J., Liu, Z., Wen, J., Fu, R., 2022. Multiple hierarchical compression for deep neural network toward intelligent bearing fault diagnosis. Engineering Applications of Artificial Intelligence 116, 105498.
[62] Van, M., Kang, H.J., 2015. Bearing defect classification based on individual wavelet local fisher discriminant analysis with particle swarm optimization. IEEE Transactions on Industrial Informatics 12, 124–135.
[63] Wang, G., Zhang, F., Cheng, B., Fang, F., 2021a. Damer: a novel diagnosis aggregation method with evidential reasoning rule for bearing fault diagnosis. Journal of Intelligent Manufacturing 32, 1–20.
[64] Wang, H., Ni, G., Chen, J., Qu, J., 2020a. Research on rolling bearing state health monitoring and life prediction based on pca and internet of things with multi-sensor. Measurement 157, 107657.
[65] Wang, Q., Wang, S., Wei, B., Chen, W., Zhang, Y., 2021b. Weighted k-nn classification method of bearings fault diagnosis with multi-dimensional sensitive features. IEEE Access 9, 45428–45440.
[66] Wang, S., Niu, P., Guo, Y., Wang, F., Li, W., Shi, H., Han, S., 2020b. Early diagnosis of bearing faults using decomposition and reconstruction stochastic resonance system. Measurement 158, 107709.
[67] Wang, X., Cui, L., Wang, H., Jiang, H., 2021c. A generalized health indicator for performance degradation assessment of rolling element bearings based on graph spectrum reconstruction and spectrum characterization. Measurement 176, 109165.
[68] Wang, X., Mao, D., Li, X., 2021d. Bearing fault diagnosis based on vibro-acoustic data fusion and 1d-cnn network. Measurement 173, 108518.
[69] Wang, Z.Y., Lu, C., Zhou, B., 2018. Fault diagnosis for rotary machinery with selective ensemble neural networks. Mechanical Systems and Signal Processing 113, 112–130.
[70] Wei, Z., Wang, Y., He, S., Bao, J., 2017. A novel intelligent method for bearing fault diagnosis based on affinity propagation clustering and adaptive feature selection. Knowledge-Based Systems 116, 1–12.
[71] Xu, Z., Li, C., Yang, Y., 2020. Fault diagnosis of rolling bearing of wind turbines based on the variational mode decomposition and deep convolutional neural networks. Applied Soft Computing 95,.
[72] Yang, Y., Wang, H., Cheng, J., Zhang, K., 2013. A fault diagnosis approach for roller bearing based on vpmcd under variable speed condition. Measurement 46, 2306–2312.
[73] Yuan, Y., Zhao, X., Fei, J., Zhao, Y., Wang, J., 2015. Study on fault diagnosis of rolling bearing based on time-frequency generalized dimension. Shock and Vibration 2015.
[74] Zhang, H., He, Q., 2020. Tacholess bearing fault detection based on adaptive impulse extraction in the time domain under fluctuant speed. Measurement Science and Technology 31, 074004.
[75] Zhang, K., Li, Y., Scarf, P., Ball, A., 2011. Feature selection for high-dimensional machinery fault diagnosis data using multiple models and radial basis function networks. Neurocomputing 74, 2941–2952.
[76] Zhang, L., Fan, Q., Lin, J., Zhang, Z., Yan, X., Li, C., 2023. A nearly end-to-end deep learning approach to fault diagnosis of wind turbine gearboxes under nonstationary conditions. Engineering Applications of Artificial Intelligence 119, 105735.
[77] Zhang, X., Liang, Y., Zhou, J., et al., 2015. A novel bearing fault diagnosis model integrated permutation entropy, ensemble empirical mode decomposition and optimized svm. Measurement 69, 164–179.
[78] Zhao, R., Yan, R., Chen, Z., Mao, K., Wang, P., Gao, R.X., 2019. Deep learning and its applications to machine health monitoring. Mechanical Systems and Signal Processing 115, 213–237.
[79] Zhao, Y., Zhang, X., Wang, J., Wu, L., Liu, Z., Wang, L., 2023. A new data fusion driven-sparse representation learning method for bearing intelligent diagnosis in small and unbalanced samples. Engineering Applications of Artificial Intelligence 117, 105513.
[80] Zheng, J., Jiang, Z., Pan, H., 2018. Sigmoid-based refined composite multiscale fuzzy entropy and t-sne based fault diagnosis approach for rolling bearing. Measurement 129, 332–342.
[81] Zheng, K., Luo, J., Zhang, Y., Li, T., Wen, J., Xiao, H., 2019. Incipient fault detection of rolling bearing using maximum autocorrelation impulse harmonic to noise deconvolution and parameter optimized fast eemd. ISA transactions 89, 256–271.
[82] Zhou, Y., Yan, S., Ren, Y., Liu, S., 2021. Rolling bearing fault diagnosis using transient-extracting transform and linear discriminant analysis. Measurement 178, 109298.
[83] Zvokelj, M., Zupan, S., Prebil, I., 2010. Multivariate and multiscale monitoring of large-size low-speed bearings using 2 ensemble empirical mode decomposition method combined with principal component analysis. Mechanical Systems and Signal Processing 24, 1049–1067.